\documentclass{article}

\usepackage{aaai}

\usepackage{graphicx}
\usepackage{subcaption}
\usepackage{lscape}
\usepackage{rotating}
\usepackage[hyphens]{url}
\usepackage{times}
\usepackage{helvet}
\usepackage{courier}
\newcommand{\citet}[1]{\citeauthor{#1}\shortcite{#1}} \newcommand{\citep}{\cite}  

\begin{document}
\title{\LARGE Is this pofma? Analysing public opinion and misinformation in a COVID-19 Telegram group chat}


\author{Ng Hui Xian Lynnette\textsuperscript{1}, Loke Jia Yuan\textsuperscript{2}\\
\textsuperscript{1}Singapore, lhuixiann@gmail.com \thanks{Any opinions, findings and conclusions or recommendations expressed in this material are those of the author and do not reflect the views of any organisation or government.}\\
\textsuperscript{2}Centre for AI and Data Governance, Singapore Management University, jyloke@smu.edu.sg\thanks{This research is supported by the National Research Foundation, Singapore under its Emerging Areas Research Projects (EARP) Funding Initiative. Any opinions, findings and conclusions or recommendations expressed in this material are those of the author(s) and do not reflect the views of National Research Foundation, Singapore.}\\
}


\maketitle              

\begin{abstract}
We analyse a Singapore-based COVID-19 Telegram group with more than 10,000 participants. First, we study the group’s opinion over time, focusing on four dimensions: participation, sentiment, topics, and psychological features. We find that engagement peaked when the Ministry of Health raised the disease alert level, but this engagement was not sustained. Second, we search for government-identified misinformation in the group. We find that government-identified misinformation is rare, and that messages discussing these pieces of misinformation express skepticism.
\end{abstract}

\section{Introduction}
\textbf{COVID-19 2020.} The novel coronavirus pandemic (COVID-19) is an ongoing global health event. In this rapidly unfolding crisis, people are unsure about what is happening and what they should do. They seek to make sense of their uncertainty \cite{sensemaking,maitlis2010sensemaking}. To do so, many people turn to new media platforms that provide support and real time information that cannot be found elsewhere \cite{stephens2009if,choi2009consumer}. Before the pandemic, more than 60\% of Singaporeans were consuming news via social media, and we expect this figure to rise as more people stay home \cite{newman2019reuters,nielsen2020}.

Public health authorities need to satisfy the public's need for information, prevent risk exaggeration, and encourage desirable behaviors like social distancing and hand-washing. Understanding online public opinion is one way to understand the efficacy of public health messaging and improve future communications. On February 15, World Health Organisation Director-General said “we’re not just fighting an epidemic, we’re fighting an infodemic”.

We study a Singapore-based Telegram group chat with more than 10,000 participants that was created to discuss COVID-19, focusing on the first six weeks of the group's existence, from 27 January to 8 March 2020. These weeks represent the ``first wave" of the pandemic in Singapore, during which the country saw the number of confirmed cases grow from 4 to 153, mostly imported from China. For two of those weeks Singapore had the most number of confirmed cases in the world outside China. During this period, the Ministry of Health raised the DORSCON (Disease Outbreak Response System Condition) level from yellow to orange. The weekly number of cases and key events in Singapore and worldwide are listed in Figure \ref{fig:cases}.

Specifically, we ask the following research questions:
\textbf{RQ1:} How does group opinion change over time?\newline
\textbf{RQ2:} How prevalent is government-identified misinformation in the group?

\textbf{Telegram public groups.} Telegram is an instant messaging service with more than 200 million monthly active users. Telegram facilitates the building of groups of up to 200,000 members. Messages in the group are only visible to people who search for or join the group, with no limits on forwarding. Telegram positions itself as a platform that protects user privacy and free expression \cite{telegramopennetwork}. Users can use the platform without revealing personally identifying information to other users. The combination of large group sizes, partial visibility and anonymity plausibly facilitates the spread of misinformation.

\textbf{Misinformation and Government Corrections.}
The Singapore government has taken steps to combat misinformation about COVID-19. The official source for updates about the local situation is a Ministry of Health web page. Prominently displayed at the top of the page are recent clarifications on misinformation. The Government has also used the Protection from Online Falsehoods and Manipulation Act (POFMA) to correct claims about COVID-19.

\section{Related Work}

\textbf{Group Chats.} Researchers have analysed the general patterns of interaction in group chats \citep{caetano_analyzing_2018,garimella2018whatapp,qiu2016lifecycle} and the use of group chats for specific purposes \cite{bouhnik2014whatsapp,wani2013efficacy}. Previous papers which study misinformation and group chats mostly do so in the context of political elections \cite{resende2019analyzing,machado2019study,10.1145/3308558.3313688}. Within the crisis informatics literature, others have studied the role of group chats in events like floods, war and kidnap responses \cite{bhuvana2019facebook,malka2015fighting,simon2016kidnapping}.

\textbf{Social media and pandemics.} \citep{liu2011organizations,wilson2018new} analyse how public health authorities use social media and other infocomm technologies for diagnostic efforts, coordination, and risk communication. Other studies focus on the public instead of health authorities: \cite{chew2010pandemics,szomszor2011twitter} study the types and source of content shared during the 2001 H1N1 outbreak; \cite{strekalova2017health} analyses audience engagement with posts from the Centers for Disease Control and Prevention (CDC) Facebook channel during the Ebola outbreak; \cite{sharma2017zika} find that misleading posts are more popular than posts containing accurate information during the Zika outbreak in the United States. 

\textbf{Contribution.} In the overall literature on the social science of new media, most studies have focused on public platforms like Facebook and Twitter, as opposed to less-visible, chat-based platforms like Telegram. Studies that analyse group chats do not focus on disease outbreaks, and studies that focus on disease outbreaks do not analyse group chats. As far as we know, our paper is the first to analyse group chats during a disease outbreak. 

\section{Methodology}
We describe how we gathered data from a public Telegram group and the methods used to process and analyse the data.

\subsection{Data collection}
Several Singapore-based public Telegram groups emerged after Singapore's first confirmed case of the coronavirus on 23 January 2020. We found the groups by searching "Singapore Coronavirus Telegram" on Google and Telegram. Some groups were: \textit{SG Fight COVID-19}\footnote{\url{http://t.me/sgVirus}}, \textit{Wuhan COVID-19 Commentary}\footnote{\url{http://t.me/WuhanCOVID}} and \textit{SG Fight Coronavirus}\footnote{\url{http://t.me/sgFight}}.

In this paper, we focus on \textit{SG Fight COVID-19} because it contains the most discussion and members. In contrast, the other groups are characterised by one-to-many news broadcasts. From 19 January to 8 March 2020, we retrieved messages with the Python Telethon API. Our method is similar to \cite{pushshift}. In total, we collected 48,050 messages. Of these, 10,765 were system-generated (e.g. automated messages about people joining and leaving the group, group name changes), leaving us with 37,285 mixed-media messages to analyse. The breakdown of messages types is presented in Table \ref{tab:stats}.

\begin{table}
\centering
\begin{tabular}{ |p{3cm}|p{1cm}| p{1.5cm}|p{1cm}|} 
 \hline
 \#Total Users & 10,765 & \#Video &  276 \\ \hline
 \#Total Messages & 48,050 & \#Audio & 36 \\  \hline
 \#Text Messages & 26,153 & \#Url & 8830\\ \hline
 \#Images & 1928 & \#Files & 62 \\ 
 \hline
\end{tabular}
\caption{Overview of \textit{sgVirus} Telegram Group}
\label{tab:stats}
\end{table}

\subsection{Data Post-Processing}
Our post-processing steps include: conversion of timestamp from UTC to GMT+8 to represent Singapore's timezone, removal of stop-words using NLTK stopword module, filtering out urls and names of news agencies (e.g. CNA, SCMP) often referred to in text messages; tokenizing text messages with NLTK's Tweet Tokenizer module\cite{10.3115/1118108.1118117}.

\subsection{Data Limitation}
Participants in our group chat may be more digitally literate compared to users of other chat platforms. While reliable demographic data about Singaporean Telegram users is not available, our personal experience is that Telegram is mostly used by people below 65 years old. People above 65 in Singapore use other messaging channels like WhatsApp. \cite{oldpeople} found that Facebook users over 65 are most likely to share misinformation. 

\section{How does group opinion change over time?}
\label{sec:analysisSec}
We analyse group opinion over time with four indicators: participation, sentiment, topics, and psychological features.

\subsection{Participation}

\begin{figure}[h]
\centering
\includegraphics[width=0.5\textwidth]{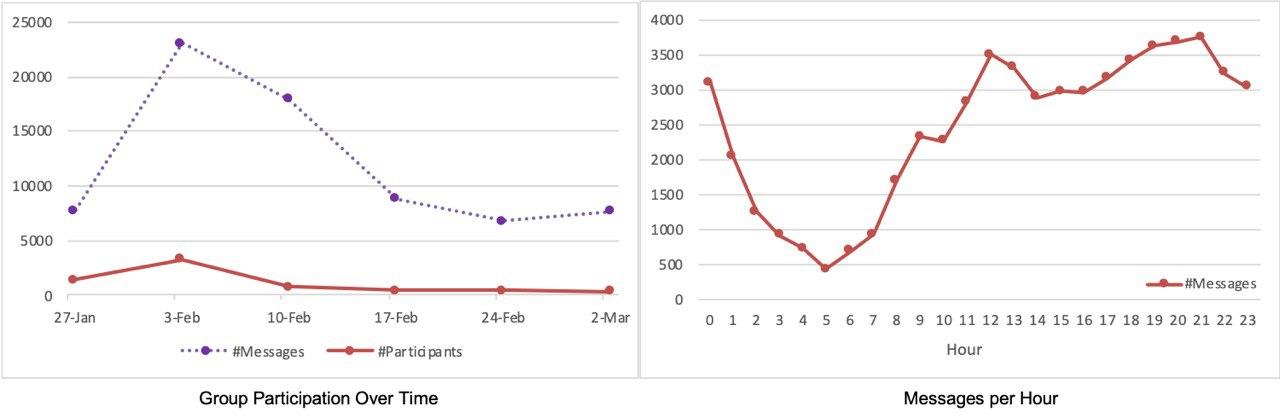}
\caption{Group Participation and Messages Over Time}
\label{fig:messages}
\end{figure}

\textbf{Active Participants.} For each week, we look at the number of users who sent at least one message, and the total number of messages. We exclude bots that forwarded messages from news websites. Our results are shown in Figure \ref{fig:messages}.

Most notable is the peak of participation in week 2. During this week, the Ministry of Health raised the disease emergency level from yellow to orange. While the news was officially announced at 5.20pm on Friday February 7, messages discussing the announcement were circulating on the group since at least 10.30am. That weekend, several supermarkets temporarily ran out of essential items and political leaders asked the public not to panic buy. After peaking in week 2, active participation fell over time. The most popular timing of messages is between 12-1pm and 8-10pm.

\textbf{Lifespan of participants.} We take the ten most active participants for each week and search for their activity in other weeks. Of the 50 most active participants, 60\% of the participants were active for 1 week only, while 40\% were active for two weeks.
 
\textbf{Discussion.} We observe that the increase in group activity corresponds with a government announcement, rather than an unusual spike in number or rate of confirmed cases. We believe this illustrates the importance of unified and coherent public health communication. The leaked information about DORSON orange prior to the official announcement likely increased uncertainty, causing people to turn to the group chat for information and support. Even true information can cause alarm, especially if it is shared in an untimely and haphazard manner. The Singapore government has recognised a need to strength internal processes and no similar leaks have occurred since \cite{leaks}.

We postulate that activity rises between 12-1pm and 8-10pm because people use their devices after lunch and dinner. Furthermore, the Ministry of Health usually releases daily updates at 8pm, triggering a flurry of discussion.

The decrease in group activity and short life span of participation suggests that the group did not meet users' needs for information. Users may have stopped relying on the Telegram group and turned to other sources. Between January-April 2020, the number of subscribers to the official government WhatsApp channel grew from 7,000 to more 900,000 \cite{sggovwhatsapp}.

\subsection{Sentiment}
To determine sentiment in the group, we perform phrase-level analysis before combining the results to obtain overall sentence-level sentiment \citep{Rezapour2018UsingLC}. This method was adopted because Telegram texts, like Tweets, are short and conversational, so traditional sentiment analysis methods for articles do not perform well. 

We first tokenized words with NLTK TweetTokenizer \cite{10.3115/1118108.1118117} which handles short expressions and strings, then identified Parts-Of-Speech(POS) tags of tokenized words with TweetNLP \cite{posonline}. Contextual sentiment was determined using the MPQA lexicon \cite{MPQA} by matching the word and the POS tag to the the appropriate entry to retrieve the corresponding contextual sentiment. The overall entry sentiment of the text message was determined through the aggregation of all the contextual sentiments.

\begin{figure}[h]
\centering
\includegraphics[width=0.5\textwidth,height=0.3\textwidth]{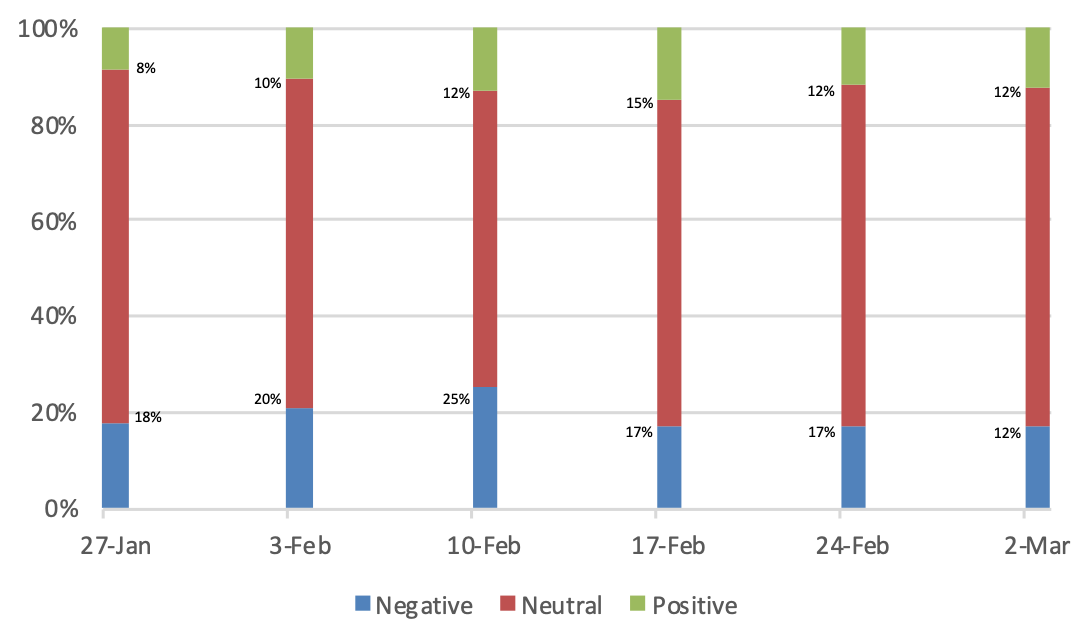}
\caption{Sentiment Dimension (Percentage) over Time}
\label{fig:sentiment}
\end{figure}

Figure \ref{fig:sentiment} represents the change in sentiment over time. We observe that there is generally more negative sentiment than positive sentiment. From Week 2 (beginning on 3 Feb) to Week 3 (beginning on 10 Feb), positive and especially negative sentiment increased. The rise in negative sentiment corresponds with the DORSCON orange weekend.

\subsection{Topics}
\begin{sidewaysfigure*}[ht]
    \includegraphics[width=1.0\textwidth]{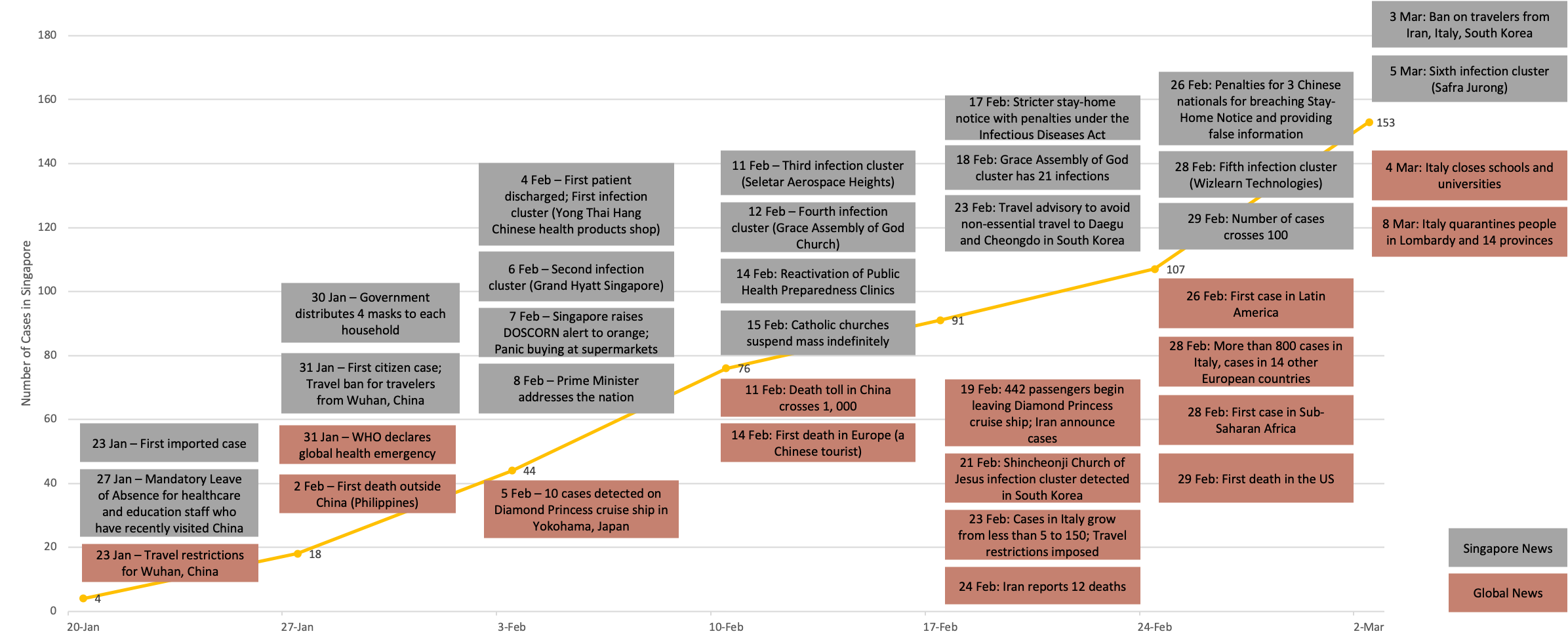}
    \caption{Timeline of major news events and number of cases in Singapore for the First Wave, where infections are from persons who come from Wuhan.\footnote{Cases were referred from The Straits Times for Singapore News and New York Times for International News. For Week 5 and 6, the situation had grown proportionally internationally, and we picked the most salient topics that the Singapore group chat was discussing.}}
    \label{fig:cases}
\end{sidewaysfigure*}

We use Mallet's Latent Dirichlet Allocation (LDA) module \cite{McCallumMALLET} to identify the top five word clusters each week. Clusters are chosen based on coherence scores and manual assessment. Where some clusters are too similar, we grouped them as one. Table \ref{tab:topics} shows the word clusters, and our interpretation of the topic associated with each cluster.  Readers can cross reference the topics with the timeline of events in Figure \ref{fig:cases}.

\begin{table}
\centering
\begin{tabular}{|p{1cm}|p{6.5cm}| } 
 \hline
  \textbf{Start Date} & \textbf{Topics \& Keywords} \\ \hline
  27 Jan & \textit{New Outbreak in China}- new, health, chinese, outbreak \newline \textit{Number of confirmed cases}- cases, confirmed, world, chinese \newline \textit{Flight and Travel restrictions} - outbreak, flight, travel, restrictions \newline \textit{Wearing masks}- masks, wear, don, think, need \\ \hline
  3 Feb &  \textit{Chinese outbreak}- outbreak, orange, travel, home, chinese, source, masks \newline  \textit{Diamond Princess Cruise Ship in Japan}- cruise, ship, japan, ncov, fake \newline 3- new, confirmed, cases, infected, quarantine \newline \textit{Don't Panic and wear masks}- masks, don, need, buy, dont wear, panic \newline \textit{Outbreak in Hong Kong}- world, outbreak, chinese, hong kong \\ \hline
  10 Feb & \textit{Wearing masks}- mask dont need wear \newline \textit{Outbreak in Hong Kong}- outbreak, world, hong kong \newline \textit{Diamond Princess cruise ship cases spike} new cases, cruise ship \newline \textit{WHO releases name of virus}- covid 19, cases, health, confirmed \newline \textit{Sourcing for masks}- masks, buy home, need source \\ \hline
  17 Feb & \textit{Grace Assembly of God church cluster}- church case, test \newline \textit{Encouraging the wearing of masks}- wear mask, dont spread, really true \newline \textit{South Korea and Italy spike in cases}- south korea, infected, italy death \newline \textit{Measures for work}- covid 19, going work safe \newline \textit{Keeping track of confirmed and discharged cases count}- confirmed case, discharged hospital, infection \\ \hline
  24 Feb & \textit{Vaccines}- flu, make vaccine, information \newline \textit{Italy cases spike}- italy death, spread \newline \textit{South Korea and Iran report cases}- covid19 cases, south korea, iran \newline \textit{Encouraging staying home}- stay home, don come work \newline \textit{Confirmation of cases}- wear mask, confirmed, moh, linked \\ \hline
  2 Mar & \textit{Wearing masks and symptoms}- wear masks, vaccine, ncov, cough, symptoms \newline \textit{Jurong SAFRA cluster}- new cases, covid 19, cluster, safra, hospital discharged \newline \textit{Cases in Iran and Italy}- cases test, iran, italy, confirmed \newline \textit{Keeping track of case counts}- infected, true cases, new patients \\ \hline
\end{tabular}
\caption{Summary of topics across the week identified using LDA}
\label{tab:topics}
\end{table}

We observe two consistently popular topics. First, topics related to ``cases". Understandably, the number of cases is a focal point in a pandemic (\textit{Feb 10:} [...]unfortunately, the ship has 60 new cases[...]; \textit{Feb 13:} A new case in NUS![...]). In all six weeks, the keyword ``cases" is almost always accompanied by ``confirmed" (e.g. week 3: cases, health, confirmed). In week 6, we observe for the first time the keyword ``true cases". Public health experts have warned that confirmed cases may not be the same as true cases, especially in countries that are under-testing \citep{warning}. We speculate that this distinction is being reflected in the group chat.

The second consistently popular topic is masks. In weeks 1 and 2, participants seem to discourage people from panicking and wearing masks (e.g. \textit{Feb 9:} Not sick dont wear mask.). In later weeks, the keyword ``masks" is no longer clustered with ``dont", suggesting that participants are starting to support mask-wearing (\textit{27 Feb:} Absolutely agreed! Wear a mask to protect ourselves and not wear when we are ill!). In parallel, we observe that participants encourage others to stay home (\textit{8 Mar: }Students stay at home. Adults try to work from home.). Our observation that participants begin to discuss socially responsible behaviors in later weeks potentially indicates that messaging from public health authorities is having an impact. 

One behavior that is not a popular topic in the group chat is hand-washing. Possible reasons include: participants do not feel that hand-washing is important, or hand-washing is an obvious and non-controversial behavior that does not warrant discussion.
 
Global events that receive attention in the group chat are the Diamond Princess Cruise ship in Japan (weeks 2 and 3), and the spread of the virus in South Korea, Italy and Iran from week 3 onward. 

\subsection{Psychological dimensions}
\begin{figure}[h]
\centering
\includegraphics[width=0.5\textwidth]{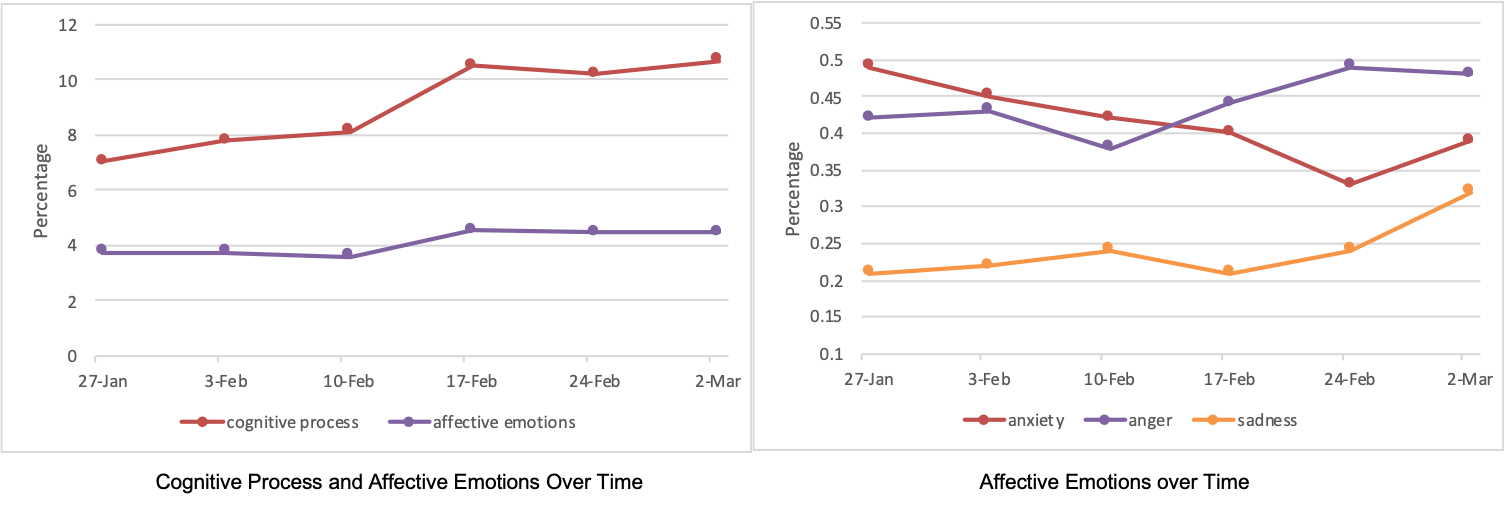}
\caption{Psychological Dimensions Over Time}
\label{fig:psycho}
\end{figure}

\textbf{Shift in emotional values.} We use the 2015 version of Linguistic Inquiry and Word Count (LIWC) \cite{doi:10.1177/0261927X09351676}. LIWC is a word-count lexicon that summarises the emotional, cognitive, and structural components in a given text sample. We focus on cognitive and affective components (Figure \ref{fig:psycho}). In terms of affective emotions, anxiety fell over time but sadness increased over time. We speculate that group chat members were becoming more certain that the pandemic is a serious event.

\textbf{Correlation between Cognitive Processes and Emotions.} We perform a Pearson correlation test on affective and cognitive dimensions (Table \ref{tab:liwc-correlations}). 

We find that positive emotions are significantly correlated with cognitive processes. We also find that anxiety has significant negative correlation with cognitive processes. By analysing 835 messages with both positive emotions and cognitive processes, we find that 20\% of messages contain words that represent hope (\textit{Feb 10: }Let’s humanity prevails and hope all good at the end of the day) and thankfulness (\textit{18 Feb: }A huge thank you to our healthcare family for providing our patients with the best care possibl [sic]). In 1148 messages with anxiety, we note many participants experience fear as this disease is unknown (\textit{18 Feb:} I never sick for more than 5 years, but still have concern. Why? As announced by the gov, there is a lot we don't know about the nCoV-19. So why take the risk?).

\begin{table}[b!]
\centering
\begin{tabular}{|p{0.8cm}|p{0.9cm}|p{0.9cm}|p{0.9cm}|p{0.8cm}|p{0.9cm}|p{0.8cm}| } 
 \hline
    & \textbf{cog-\newline proc} & \textbf{in-\newline sight} & \textbf{ten-\newline tat} & \textbf{cer-\newline tain} & \textbf{cause} & \textbf{dis-\newline crep} \\ \hline
    \textbf{affect} & \textbf{0.94} & \textbf{0.92} & \textbf{0.95} & \textbf{0.90} & \textbf{0.91} & \textbf{0.99} \\ \hline
    \textbf{pos-\newline emo} & \textbf{0.91} & \textbf{0.91} & \textbf{0.95} &0.80 & \textbf{0.86} & \textbf{0.88} \\ \hline
    \textbf{neg-\newline emo} & 0.58 & 0.52 & 0.49 & 0.76 & 0.62 & 0.54 \\ \hline
    \textbf{anx} & \textbf{-0.83} & \textbf{-0.82} & \textbf{-0.86} & -0.70 & \textbf{-0.87} & -0.73 \\ \hline
    \textbf{anger} & 0.72 & 0.66 & 0.72 & 0.74 & 0.75 & 0.51 \\ \hline
    \textbf{sad} & 0.53 & 0.50 & 0.42 & 0.71 & 0.58 & 0.51 \\
 \hline
\end{tabular}
\caption{Correlation values between Cognitive Process (rows) and Affective (columns) dimensions. Statistically significant values where $p<0.05$ are highlighted in \textbf{bold}}
\label{tab:liwc-correlations}
\end{table}

\begin{figure}[h]
\centering
\includegraphics[width=0.5\textwidth]{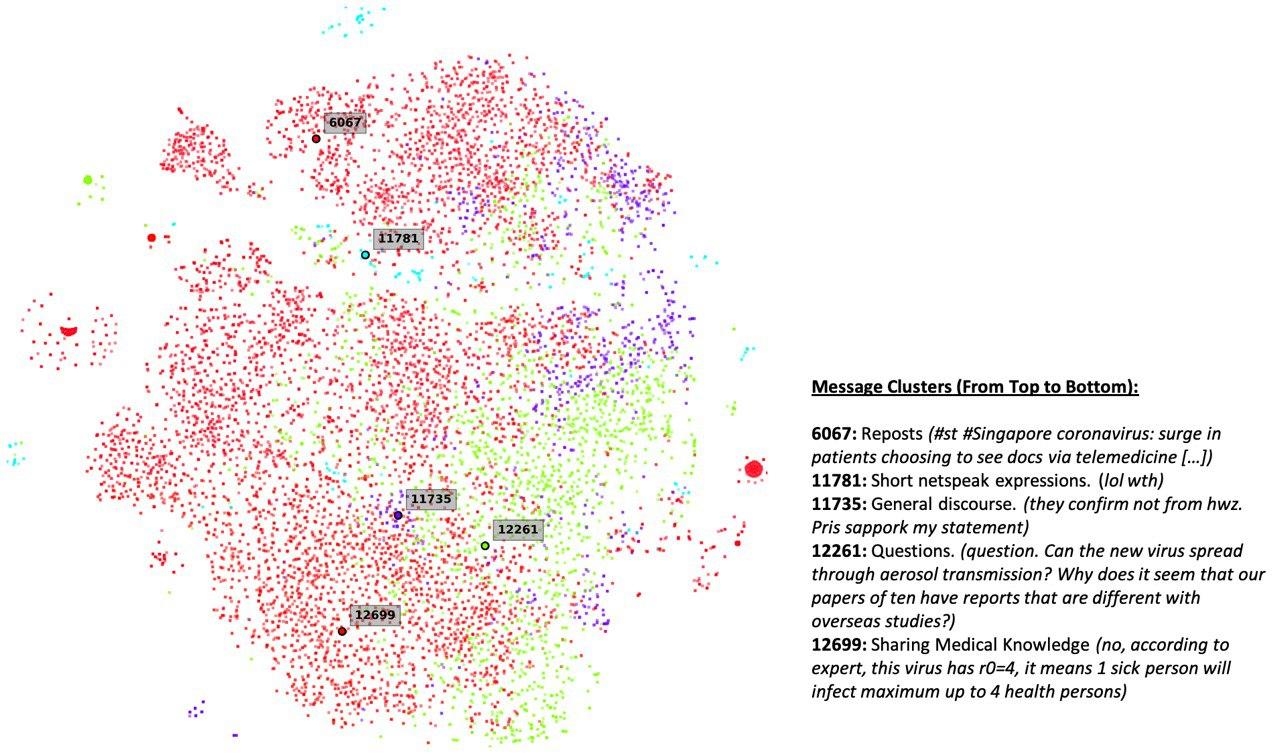}
\caption{Clusters of Text Messages}
\label{fig:tsne}
\end{figure}

\textbf{Visualising LIWC Features.} We reduce our n-dimensional LIWC feature set into a 2D space using Singular Value Decomposition (SVD). We first perform SVD by using different singular value numbers from k=0 to k=50, and finally select k=20 by the use of the elbow rule heuristic on the singular values generated across the different k-values. We visualise the salient set of compressed features with a t-SNE plot, and use a k-Nearest-Neighbours clustering algorithm to find cluster means, visualised in Figure \ref{fig:tsne}. In this cluster analysis of text messages using linguistic features, we observe that users participate in five main clusters: (1) Reposts from news websites, (2) Short netspeak expressions, (3) General discourse, (4) Questions and (5) Sharing medical knowledge.

\section{How prevalent is government-identified misinformation?}
We analyse the prevalence of misinformation in the group chat, focusing on pieces of misinformation which have been corrected by the Singapore government. We discuss participants' reactions to these pieces of misinformation.

\subsection{Identifying misinformation}
 We refer to the list of corrections about COVID-19 provided by the Ministry of Health (MOH)\footnote{\url{https://www.moh.gov.sg/covid-19/clarifications}} and the government's fact-checking web page Factually\footnote{\url{https://www.gov.sg/factually}}. Between 24 January and 8 March 2020, the two sites listed 17 corrections. The fact that the Singapore government has addressed these pieces of misinformation suggests that they have been identified as particularly harmful. 

Our approach (relying on a list created by a fact-checking third party) is borrowed from others including \cite{10.1145/3308558.3313688,resende2019analyzing}. We recognise that the third-party (in our case, the Singapore government) may not have identified every piece of misinformation. This limitation speaks to wider challenges in the study of misinformation, namely the difficulty of establishing a ground-truth standard of what constitutes misinformation.

To search for misinformation in the group chat, we focus on the five days surrounding each piece of misinformation: two days before, the day of clarification, and two days after. For textual messages, we first filter for keywords, then manually screen the results. For example, to identify messages challenging the ``validity of the maskgowhere.gov.sg site", we perform automatic keyword filtering for messages containing ``maskgowhere", which returns 6 results. However, all 6 messages discuss the site in general instead of challenging the site's validity, so our final result is 0. For images, videos, audio and urls, we perform a manual search. 

\subsection{Results}
We find that government-identified misinformation is rare on the group chat. 6 (out of 17) pieces of misinformation are discussed in the group chat. These pieces of misinformation are contained in 18 textual messages, 3 urls, 6 images and 1 video, representing 0.05\% of all messages. Full results are shown in Table \ref{tab:rq2}. Moreover, we find that messages that discuss misinformation tend to express skepticism. Participants seek to verify the accuracy of content rather than simply ``passing it on".(\textit{12 Feb:} jolibee (sic) lucky plaza got case?).

Participants' attitudes towards misinformation deserve a comprehensive investigation, but here we report two preliminary, interesting observations. First, 0.4\% of all messages contain ``Is this fake" or variations (``true or fake", ``true or not"), almost 10x more than messages discussing misinformation. This suggests that participants are aware of the prevalence of misinformation. Many people are not passive consumers; they seek to actively check multiple sources and verify information \citep{dubois2018echo}.

Second, in 63 instances, participants use ``POFMA" in new ways. The POFMA (Protection from Online Falsehoods and Manipulation Act) is a Singapore law passed in 2019 aimed at correcting misinformation. Originally the name of a law, the term is being used as a verb (``POFMA me), an entity with agency (``POFMA busy), and a byword for misinformation (``that doesn't look like censoring but pofma").

\begin{table}
\centering
\begin{tabular}{ |p{1cm}|p{2cm}|p{2cm}|p{2cm}| } 
 \hline
    \textbf{Date} & \textbf{Description} & \textbf{Keywords Matches} & \textbf{Sample Messages} \\ \hline 
    14 Feb & False statements on States Times Review Facebook page & 2 texts: States Times Review (2) \newline 1 url & SST is known for publishing fake news \\ \hline
    13 Feb & Graphic aired on CNA Asia listed a death to Singapore instead of Hong Kong & 3 texts: CNA (3), death(3) \newline 5 images, 2 url & Antonio: Isn't the 1 death from Philippines not Singapore? \newline Is this true? Can't find other source for the 1 death in sg \\ \hline
    12 Feb & Voice recording via text messaging platforms advising persons to avoid Lucky Plaza after an individual had fainted at Jollibee & 2 texts: lucky plaza (2), jollibee (2) & jolibee (sic) lucky plaza got case? \newline Heard theres a lock down now on the Sixth Floor of Lucky Plaza. Who can verify? \\ \hline
    7 Feb & Message on a death in Singapore due to the virus & 5 texts: died (5) & Singapore dorscon level will change to orange at 2.30pm. The condition of 2 people with the virus have worsened. 1 of them in critical condition. \newline most likely announce orange follow by 1 death. \\ \hline
    28 Jan & Woodlands MRT was closed for disinfection & 4 texts: Woodlands mrt (1), disinfection (4) & is woodlands mrt really closed for disinfection? \\ \hline
    24 Jan & Suspected case at EastPoint Mall, who visited raffles medical center & 4 texts: Eastpoint mall(4), raffles medical (1) \newline 1 image \newline 1 video & is this verified? (regarding eastpoint suspected case) \\
 \hline
\end{tabular}
\caption{Government Clarifications and Chatter on \textit{sgVirus} group}
\label{tab:rq2}
\end{table}

\section{Conclusion}
We analyse a Telegram group chat about COVID-19 in Singapore. The group was most active from 3-9 February. This week is notable because of a leaked press release prior to the official shift to DORSCON orange. We believe this demonstrates the importance of coherent and unified public health communication. User participation is short-lived, plausibly indicating that the group chat did not meet users' needs for information and support. Across the weeks, emotions shifted from anxiety to sadness, and negative sentiment decreased. 

We also find that government-identified misinformation is rare on the group chat. Messages that discuss these pieces of misinformation tend to be skeptical. In general, participants often express doubt about the validity of content; they are not passive consumers of (mis)information.

Our study is a preliminary attempt to investigate an ongoing and dynamic crisis. It contributes to the small but growing literature on group chats as platforms for sharing (mis)information. A single Telegram group is not representative of the rest of the population, so it is hard to tell if the findings are generalizable. Further research can include more group chats, compare cross platform group chats, extend the analysis to cover the ``second wave" in Singapore, and/or look for other types of misinformation, not just government-identified misinformation. 

{\small \bibliography{ref.bib}}

\bibliographystyle{aaai}

\end{document}